\documentclass[12pt]{iopart}

\newcommand{\al}{\alpha}
\newcommand{\be}{\beta}
\newcommand{\ga}{\gamma}

\newcommand{\de}{\delta}

\newcommand{\m}{\mu}
\newcommand{\n}{\nu}
\newcommand{\rh}{\rho}
\newcommand{\si}{\sigma}
\newcommand{\Si}{\Sigma}
\newcommand{\ta}{\tau}
\newcommand{\om}{\omega}
\newcommand{\mc}{\mathcal}

\newcommand{\pie}{\mbox{\Large$\pi$}}

\newcommand{\ce}{\mc{E}}
\newcommand{\cs}{\mc{S}}
\newcommand{\cl}{\mc{L}}
\newcommand{\ck}{\mc{K}}
\newcommand{\cb}{\mc{B}}

\newcommand{\ca}{\mc{A}}

\newcommand{\p}{\perp}
\newcommand{\pl}{\partial}
\newcommand{\na}{\nabla}
\newcommand{\nd}{\dot{n}}

\usepackage{amsbsy,amssymb}

\begin{document}
\title[Covariant Schwarzschild perturbations I]{Covariant Schwarzschild perturbations I: Initial value formulation for scalars of spin-weight $\pm 2$}
\author{R. B. Burston and A. W. C. Lun}
\address{School of Mathematical Sciences, Monash University, Australia}
\ead{Raymond.Burston@gmail.com}
\ead{Anthony.Lun@sci.monash.edu.au}
\begin{abstract}
We consider full perturbations to a covariantly defined Schwarzschild spacetime. By constructing complex quantities, we derive two decoupled, covariant and gauge-invariant, wave-like equations for spin-weighted scalars. These  arise naturally from the Bianchi identities and comprise a  covariant representation of the Bardeen-Press equations for scalars with spin-weight $\pm2$. Furthermore, the covariant and gauge-invariant 1+1+2 formalism is employed, and consequently, the physical interpretation of the energy-momentum perturbations is transparent. They are written explicitly in terms of the energy-momentum specified on spacelike three-slices. Ultimately, a Cauchy problem is constructed whereby, an initial three-slice may be perturbed by an energy-momentum source, which induces resultant gravitational fields. 
\end{abstract}
\maketitle
\section{Introduction}

In this paper, we consider perturbations to a background Schwarzschild spacetime which is described covariantly in terms of three non-vanishing locally rotationally symmetric (LRS) class II scalars \cite{Clarkson}.   Using the LRS class II symmetries, the background may be foliated with a family of spacelike three-slices \cite{Ellis1973}, and every three-slice is further foliated by a family of two-slices. The energy-momentum is specified as a first-order perturbation to an initial spacelike three-slice, and a Cauchy problem is constructed.  This energy-momentum is what drives, or induces, any subsequent perturbed gravitational fields. A interesting example can be constructed by specifying a particular astrophysical object as the source to any resultant gravitational radiation.

We use the 1+1+2 system of equations governing the first-order geometric and kinematic quantities which arise from both the Bianchi and Ricci identities. In this paper, we generalize the system given in \cite{Clarkson} to include energy-momentum sources.

We then demonstrate how to derive covaraint wave-like equations arising from the once-contracted Bianchi identities. Upon inspecting the fully non-linear equations, we note that the successful decoupling method described in \cite{BurstonEM1} is also suitable for the system that arises here. Thus we construct complex tensors for the irreducible parts of the gravito-electromagentic (GEM) fields. Subsequently, by using a covariant decomposition of the two-sheet described in \cite{BurstonEM1}, we derive two decoupled wave-like equations for complex spin-weighted \cite{Penrose} scalars. These are a covariant representation of the Bardeen-Press (BP) equations \cite{Bardeen}. Furthermore, the energy-momentum source terms have been included to allow pragmatic calculations of perturbed gravitational fields arising from initial energy-momentum perturbations.

\section{Decomposing the Spacetime}

The necessary mathematical tools for decomposing tensors are separated into three distinct parts.  The first represents the {\it 1+3 formalism} which is well established throughout the literature (for example, see \cite{Bel1958,Ehlers1993,Ellis1967}). The second focuses on a further decomposition using the recently developed {\it 1+1+2 covariant two-sheet formalism} \cite{Clarkson}. Finally, a further {\it covariant} decomposition of the two-sheet, expanding the work of \cite{BurstonEM1}, is described. As this paper uses the methods developed in \cite{BurstonEM1},  we retain that notation throughout this manuscript\footnote{The transformation between the notation used in this manuscript to that of \cite{Ehlers1993,Ellis1967,Ellis1973} is \mbox{$(n^\m,\perp_{\m\n},A_{\m\n},K,\rho,j_\m,B_{\m\n})\rightarrow(u^\m,h_{\m\n},-\sigma_{\m\n},-\Theta,\mu,q_\m,-H_{\m\n}).$}}.

\subsection{1+3 Decomposition}

The tools required to irreducibly decompose the equations governing general relativity into a covariant 1+3 form are presented here.  An observer is moving in the direction of a timelike vector ($n^\m$) which is normalized such that
\begin{eqnarray}
g^{\al\be} n_\al n_\be    : = -1.
\end{eqnarray}
The projection tensor is used to project any quantity onto the instantaneous rest space of the observer \cite{Ellis1967} and is therefore defined,
\begin{eqnarray}
\p_{\m\n}: = g_{\m\n} +n_\m n_\n.\label{projection_tensor}
\end{eqnarray}
Let $V_{\m\n}$ represent any tensor, then the {\it projected, symmetric and trace-free} (PSTF) part of $V_{\m\n}$ is defined using angular brackets,
\begin{eqnarray} 
V_{<\m\n>} :=  \p \,V_{(\m\n)} -\frac 13\,\perp_{\m\n} \perp^{\al\be} V_{\al\be},
\end{eqnarray}
where $\p$ represents an operator which projects all {\it free} indices, and the standard symmetric (round) brackets have been employed. A spacelike tensor, otherwise referred to as a {\it three-tensor}, will vanish when any of its indices are contracted with $n^\m$.

The fundamental geometric quantities are found from the decomposition of the covariant derivative of the timelike vector,
\begin{eqnarray}
 \na_\m n_\n = -A_{\m\n}-\frac13\,\p_{\m\n}\,K -n_\m \nd_\n +{\epsilon_{\m\n}}^\be \omega_\be.
\end{eqnarray}
Here, $K:= -\perp^{\al\be} \na_\al n_\be$ is the {\it expansion} and  $A_{\m\n}:= -\na_{<\m} n_{\n>}$ is the shear. The spacelike acceleration is denoted $\nd_\m := n^\al \na_\al n_\m$, and in general a ¨dot¨ derivative of any tensor is $\dot V_{\m\dots\n} := n^\al \na_\al V_{\m\dots\n}$. The spacelike vorticity is $\omega_\m:= \frac 12\, {\epsilon_\m}^{\al\be} \na_\al n_\be$, and the completely anti-symmetric three-Levi-Civita psuedo-tensor is defined such that $\epsilon_{\m\n\si}:=\epsilon_{\al\m\n\si}n^\al$, where  $\epsilon_{\m\n\si\ta}$ is the usual Levi-Civita pseudo tensor.

The trace-free part of the Riemann tensor, otherwise know as the Weyl conformal curvature tensor ($C_{\m\n\si\ta}$), can be expressed in terms of two spacelike, trace-free gravito-electromagnetic (GEM) tensors \cite{Ellis1973},
\begin{eqnarray}
E_{\m\n}:= C_{\al\m\be\n}n^\al n^\be \qquad\mbox{and}\qquad B_{\m\n}:=\frac 12 \epsilon_{\al\m\be\gamma} {C^{\be\gamma}}_{\om\n} n^\al n^\om.
\end{eqnarray}
Here $E_{\m\n}$ is the gravito-electric field and $B_{\m\n}$ the gravito-magnetic field.

Finally, the energy-momentum tensor ($T_{\m\n}$) may also be irreducibly decomposed into 1+3 form,
\begin{eqnarray}
\fl\rho:= T_{\al\be}n^\al n^\be ,\,\,\,\,\,\,\, P:= \frac 13 \p^{\al\be} T_{\al\be}, \,\,\, \,\,\,\,j_\m :=-\p\,T_{\m\al}n^\al \qquad\mbox{and} \qquad
\pie_{\m\n}:=T_{<\m\n>}\label{eq2} .
\end{eqnarray}
Here $\rh$ is the mass-energy density, $P$ is the isotropic pressure, $j_\m$ is the spacelike mass-energy flux and $\pie_{\m\n}$ represents the spacelike, trace-free, anisotropic pressure. 

Before proceeding to the next decomposition, it is important to define two primary derivatives used throughout. The Lie derivative in the direction of a vector field $X^\m$ is denoted $\cl_X $.  Let any tensor be  denoted ${V_{\m\dots\n}}^{\si\dots\ta}$ and therefore
\begin{eqnarray}
 \mc{L}_X  {V_{\m\dots\n}}^{\si\dots\ta} :=& X^\al \na_\al {V_{\m\dots\n}}^{\si\dots\ta}    \nonumber \\
  &-  {V_{\m\dots\n}}^{\al\dots\ta}    \na_\al X^\si-\dots -   {V_{\m\dots\n}}^{\si\dots\al}    \na_\al X^\ta \nonumber \\
                               & +  {V_{\al\dots\n}}^{\si\dots\ta} \na_\m X^\al+\dots+   {V_{\m\dots\al}}^{\si\dots\ta}   \na_\n X^\al .
\end{eqnarray}
In the case of a scalar ($\varphi$) this reduces to a directional derivative
\begin{eqnarray}
\cl_X \varphi := X^\al \na_\al\varphi.
\end{eqnarray}
It is noted that the Lie derivative of a {\it covariant, spacelike tensor} is also spacelike.

The covariant three-derivative ($D_\m$) has the restriction that it operate on spacelike quantities only. Let ${W_{\m\dots\n}}^{\si\dots\ta}$ represent any three-tensor $({W_{\m\dots\n}}^{\si\dots\ta} n^\m=0,\,\dots,{W_{\m\dots\n}}^{\si\dots\ta}n_\ta=0$), then the covariant three-derivative is defined
\begin{eqnarray}
D_\ga {W_{\m\dots\n}}^{\si\dots\ta} := \p \na_\ga{W_{\m\dots\n}}^{\si\dots\ta} .
\end{eqnarray}

\subsection{1+1+2 Decomposition}

The further decomposition of the 1+3 quantities into 1+1+2 form is now presented. This is a {\it covariant, and gauge-invariant}, decomposition  recently developed by \cite{Clarkson}, however, we again adhere to the notations and conventions employed in \cite{BurstonEM1}\footnote{Where different, the transformation from the notation used in this paper to that of \cite{Clarkson} is $\{\ck,N^\m,\cs_{\m\n},K,\Si,\Si_\m ,\Si_{\m\n},\ca_{\m\n},\cb, \cb_\m,\cb_{\m\n},\hat N_\m\}$ $\rightarrow\{-\phi,n^\m,N_{\m\n},-\Theta,-\Si,-\Si_\m ,-\Si_{\m\n},-\zeta_{\m\n},-\mc{H},-\mc{H}_\m,-\mc{H}_{\m\n},a_\m\}$.}. Another vector is defined $(N^\m)$ which is both spacelike ($N_\al n^\al=0$) and normalized according to,
\begin{eqnarray}
 \perp^{\al\be}\,N_\al N_\be = 1.
\end{eqnarray}
Therefore, another projection tensor that now projects all quantities (orthogonal to both $N^\m$ and $n^\m$) onto a two-sheet \cite{Clarkson} is,
\begin{eqnarray}
\mc{S}_{\m\n}:= \p_{\m\n} - N_\m\, N_\n .
\end{eqnarray}
Furthermore, following the work of \cite{Clarkson}, the PSTF part of some tensor $(V_{\m\n})$, with respect to $N^\al$, is defined
\begin{eqnarray}
V_{\{\m \n\}} :=  V_{(\bar\m\bar\n)} -\frac 12 \, \cs_{\m\n} \,\cs^{\al\be} V_{\al\be},
\end{eqnarray}
where a ``bar" over an index implies that the index is projected onto the two-sheet, i.e.
\begin{eqnarray}
V_{\bar\m \bar \n} := {\cs_\m}^\al {\cs_\n}^\be \, V_{\al\be}.
\end{eqnarray}
A tensor which vanishes when any index is contracted with both $n^\m$ and $N^\m$ is referred to a {\it two-tensor}.
 
The important geometric quantities arise from the irreducible 1+1+2 decomposition of the covariant three-derivative of $N^\m$, 
\begin{eqnarray}
D_\m N_\n = -\mc{A}_{\m\n} -\frac 12 \cs_{\m\n} \,\mc{K}+N_\m \hat{N}_\n +\xi\,\epsilon_{\al\be}.
\end{eqnarray}
 The expansion of the two-sheet is $\ck:= -\cs^{\al\be} D_\al N_\be$. The shear of the two-sheet is $\mc{A}_{\m\n} := - D_{\{\m} N_{\n\}}$, which is symmetric and {\it trace-free with respect to the two-sheet}, i.e. $ \cs^{\al\be} \ca_{\al\be}=0$. We also denote $\hat {N}_\m:=N^\al D_\al N_\m$, and in general the ``hat'' derivative of any three-tensor (${W_{\m\dots\n}}^{\si\dots\ta}$) is defined $ {{\hat {W}}_{\m\dots\n}}^{\qquad\si\dots\ta} := N^\al D_\al {W_{\m\dots\n}}^{\si\dots\ta}$. The scalar $\xi:= \frac 12\,\epsilon^{\al\be} D_\al N_\be$ represents a twisting of the two-sheet, and the anti-symmetric two-Levi-Civita tensor is defined such that $\epsilon_{\m\n} :=  \epsilon_{\al\m\n} N^\al$.

As expressed in \cite{Clarkson}, the 1+3 geometric and Weyl quantities are decomposed into the irreducible set $
\{K,\ca,\Omega,\Si,\ce,\cb,\ca_\m,\Omega_\m,\Si_\m ,\ce_\m,\cb_\m,\Si_{\m\n}, \ce_{\m\n},\cb_{\m\n}\}$ according to
\begin{eqnarray}
\dot N_\m = \alpha_\m+\ca \, n_\m,\\
\nd_\m =\ca_\m + \ca\,N_\m,\\
\omega_\m = \Omega_\m+\Omega\, N_\m,\\
A_{\m\n}= \Si_{\m\n}-\frac12\, S_{\m\n} \Si +2\, \Si_{(\m} N_{\n)}+\Si\,N_\m N_\n, \label{1p1p2decoforamu}\\
E_{\m\n}= \ce_{\m\n}-\frac12\, S_{\m\n} \ce +2\, \ce_{(\m} N_{\n)}+\ce\,N_\m N_\n, \\
B_{\m\n}= \cb_{\m\n}-\frac12\, S_{\m\n} \cb +2\, \cb_{(\m} N_{\n)}+\cb\,N_\m N_\n.
\end{eqnarray}
Here  $\ca := -n^\al \dot N_\al = N^\al \dot n_\al$,  $\alpha_\m := \dot N_{\bar\m}$, $\ca_\m := \nd_{\bar\m}$,  $\Omega_\m := \omega_{\bar\m}$ and  $\Omega:= N^\al \omega_\al$. Furthermore, $\Si:= A_{\al\be}N^{\al }N^\be$, $\Si_\m:= A_{\bar\m\al} N^\al$ and $\Si_{\m\n}:=A_{\{\m\n\}}$, and now $\Si_\m$ and $\Si_{\m\n}$ are both two-tensors. Analogous definitions follow for both $E_{\m\n}$ and $B_{\m\n}$.

The 1+3 energy-momentum quantities, in  1+1+2 form, are
\begin{eqnarray}
j_\m = \mc{J}_\m+ \mc{J}\, N_\m,\\
\pie_{\m\n}= \Pi_{\m\n}-\frac12\, S_{\m\n} \Pi +2\, \Pi_{(\m} N_{\n)}+\Pi\,N_\m N_\n,
\end{eqnarray}
where $\mc{J}_\m := j_{\bar\m}$, $\mc{J} := N^\al j_\al$ and the decomposition for $\pie_{\m\n}$ is analogous to \eref{1p1p2decoforamu}. 
Finally, the covariant two-derivative associated with the two-sheet ($d_\m$) must operate on two-tensors only. Let ${U_{\m\dots\n}}^{\si\dots\ta}$ represent any two-tensor, then
\begin{eqnarray}
d_\ga {U_{\m\dots\n}}^{\si\dots\ta}  := D_{\bar \ga} {U_{\bar\m\dots\bar\n}}^{\bar\si\dots\bar\ta}  .
\end{eqnarray}

\subsection{Decomposing the Two-Sheet}

This final section gives a fuller description of the covariant decomposition of the two-sheet as presented in \cite{BurstonEM1}.  The decomposition of the two-sheet which is suitable for decoupling the Bianchi identities follows a different trend as the 1+3 and 1+1+2 decompositions.  A natural decomposition of the two-sheet may be expressed by defining a complex-conjugate pair of vectors, $m^\m$ and ${\bar{m}}^\m$, which satisfy the following relationships \cite{Penrose}:
\begin{eqnarray}
{\bar m}^\al m_\al =1,\qquad m^\al m_\al =0, \qquad {\bar m^\al }{\bar m_\al} =0.\end{eqnarray}
This complex-conjugate pair are orthogonal to both $n^\m$ and $N^\m$. Consequently, they can be raised and lowered using $\cs^{\m\n}$.
The projection tensor associated with the two-sheet may now be covariantly decomposed according to
\begin{eqnarray}
 \mc{S}_{\m\n} = 2\, m_{(\m} \,{\bar m}_{\n)}.\label{mbarprops}
\end{eqnarray}

Consider the arbitrary two-tensors $U_\m$ and $U_{\m\n}$, i.e. a contraction of any of their indices with $n^\m$ or $N^\m$ will vanish. By using \eref{mbarprops}, and recalling that $U_{\m\n}$ can always be split into a symmetric plus an anti-symmetric part, they are irreducibly decomposed according to
\begin{eqnarray}
U_\m = (U_\al m^\al) \bar m_\m + ( U_\al \bar m^\al) m_\m,\label{deompofum}\\
 U_{\m\n} = ( U_{\al\be} \bar m^\al \bar m^\be)  m_\m  m_\n+( U_{\al\be} m^\al m^\be) \bar m_\m \bar m_\n+ 2\, m_{(\m} \bar m_{\n)}\,(U_{\al\be} \bar m^{(\al} m^{\be)})\nonumber\\
+2\, m_{[\m} \bar m_{\n]}\,(U_{\al\be} \bar m^{[\al} m^{\be]}) .\label{antisyshedec}
\end{eqnarray}
The generalization to tensors of more complicated type is clear.

We now discuss the concept of spin-weighted objects \cite{Penrose}. Let the complex null vector, $m^\m$, undergo a transformation on the two-sheet according to 
\begin{eqnarray}
m^\m \rightarrow \mc{C} \,\bar{\mc{ C}}^{-1}\, m^\m\label{spinedef},
\end{eqnarray}
where $\mc{C}$ is an arbitrary complex scalar field and $\bar \mc{C}$ its complex conjugate. Then any quantity, ${\zeta_{\m\dots\n}}^{\si\dots\ta}$,  which has a corresponding transformation of
\begin{eqnarray}
{\zeta_{\m\dots\n}}^{\si\dots\ta}\rightarrow \mc{C}^p\, \bar \mc{C} ^q\,{\zeta_{\m\dots\n}}^{\si\dots\ta},
\end{eqnarray}
is said to have a spin-weight $s$ defined\footnote{The boost weight $(b)$ can also be defined,  $b=\frac 12\,(p+q)$.},
\begin{equation}
s:= \frac 12\,(p-q).
\end{equation}
Let us now reconsider the decomposition of $U_\m$ according to \eref{deompofum}. Since this is a two-tensor, there are only two independent scalars arising, namely  $(U_\al m^\al)$ and $(U_\al \bar m^\al)$. Under the transformation \eref{spinedef}, we have
\begin{eqnarray}
(U_\al m^\al)\rightarrow \mc{C} \,\bar{\mc{ C}}^{-1}\,(U_\al m^\al),
\end{eqnarray}
thus $(p,q)=(1,-1)$ and it therefore has a spin-weight of $s=1$. An analogous consideration shows that $(U_\al \bar m^\al) $ has spin-weight $s=-1$. Similarly, in the decomposition of $U_{\m\n}$, there are four independent scalars i.e. $U_{\al\be} m^\al m^\be$, $U_{\al\be} \bar m^\al \bar m^\be$, $\,U_{\al\be} \bar m^\al  m^\be$ and  $U_{\al\be} m^\al \bar m^\be$, which have respective spin-weights of $s=2$, $s=-2$, $s=0$ and $s=0$. Thus one must be aware of how the transformation properties of spin-weight scalars differ from the usual zero spin-weighted scalars.

We now consider the important quantities which arise from the decomposition of the covariant two-derivative of the complex-conjugate vectors. All the subsequent equations will naturally occur in conjugate pairs. However, we only display one of the pair and note that the other is found by taking the complex conjugate. Thus we have
\begin{eqnarray}
d_\m m_\n = -m_\m m_\n \bar m^\al \chi_\al -\frac 12\,\cs_{\m\n} m^\al \chi_\al -\rmi\,\frac 12 \,m^\al \chi_\al \,\si\,\epsilon_{\m\n} \label{sheedecopdodm},
\end{eqnarray}
where
\begin{eqnarray}
 \chi_\m := m^\al d_\m\bar m_\al=-\bar \chi_\m
\end{eqnarray}
 is purely imaginary and has zero spin-weight. We also define
\begin{eqnarray}
\si:=\rmi\, m_\al \bar m_\be \epsilon^{\al\be} \qquad\mbox{from which it follows}\qquad\si^2=1.
\end{eqnarray}
Thus $\si$ has zero spin-weight and since $\si=\bar\si$, it is purely real.

Therefore, \eref{sheedecopdodm} depends only on the complex-conjugate pair $(m^\al \chi_\al,\bar m^\al \bar\chi_\al)$ which have spin-weights $1$ and $-1$ respectively. We also have a constraint and a relationship for the divergence, which are respectively
\begin{eqnarray}
(d^\al+\chi^\al )m_\al =0\qquad\mbox{and}\qquad\epsilon^{\al\be} d_\al m_\be= \rmi\,\si\, d^\al m_\al.
\end{eqnarray}
Finally, using \eref{antisyshedec} the two-Levi-Civita tensor is decomposed as
\begin{eqnarray}
\epsilon_{\m\n}=\rmi\,2\,\si\,m_{[\m} \bar m_{\n]}.
\end{eqnarray}

\section{Background Schwarzschild Spacetime}

In the background spacetime, all the vorticity terms vanish, i.e. $\Omega=0$ and $\Omega_\m=0$. Therefore, {\it under the conditions of vanishing vorticity only}, the timelike vector $n^\m$ is now normal to a spacelike three-slices and according to Frobenius´ theorem, this implies $n^\m$ must satisfy
\begin{eqnarray}
n_{[\m} \na_\n n_{\si]}=0.\label{frobwed}
\end{eqnarray}
The most general solution to \eref{frobwed} is
\begin{eqnarray}
n_\m = -\al \, \na_\m f,
\end{eqnarray}
where $\al$ is the lapse function and $f$ is the Cauchy time function such that $\al\na_\m f >0$. We now introduce two more vectors: $t^\m$ which is tangent to the temporal coordinates world line and a spacelike shift vector $\be^\m:=\p t^\m$. The tangent is decomposed  according to
\begin{eqnarray}
t^\m =\be^\m +(\al\,\cl_t f)\, n^\m.
\end{eqnarray}
Therefore, the Lie derivative (in the directon of $n^\m$) operating on a {\it covariant and spacelike} tensor becomes,
\begin{eqnarray}
\cl_n = \frac 1{\al\,\cl_t f}\,(\cl_t -\cl_\be).\label{1p4lide}
\end{eqnarray}
Similarly, the twisiting of the two-sheet in the background also vanishes i.e. $\xi=0$. Therefore, {\it under the additional condition of zero twisting}, the two-sheet meshes to form a two-surface, where $N^\m$ is the corresponding normal. Since we require this normal to be hypersurface orthognal, we again use Frobenius´ theorem to conclude
\begin{eqnarray}
N_\m = \ga \, D_\m h,
\end{eqnarray}
where $\ga$ and $h$ are analogous to the lapse and Cauchy time functions such that $\ga \, D_\m h>0$. A spacelike vector, $r^\m$, is defined to be tangent to the radial coordinates worldline and a shift vector for the two-surface is defined $b^\m := r^{\bar\m}$. Then $r^\m$ is decomposed according to
\begin{eqnarray}
r^\m = b^\m+ (\ga\,\cl_r h )\, N^\m. 
\end{eqnarray}
 Finally, the Lie derivative (in the direction of $N^\m$), when operating on a covariant two-tensor becomes
\begin{eqnarray}
\cl_N = \frac 1{\ga\,\cl_r h}\,(\cl_r -\cl_b).\label{cldecomsa}
\end{eqnarray}
Hitherto, \eref{frobwed}-\eref{cldecomsa} have only assumed that the background spacetime have vanishing vorticity and twisting. They constitute a covariant description of how the Lie derivatives and normal vectors are aligned with respect to coordinate tangents in vorticity/twist free spacetimes.

We now impose further restrictions to arrive at a covariant description of the Schwarzschild spacetime. It was shown in \cite{Clarkson,Elst1996} that the background Schwarzschild spacetime may be covariantly described by only three non-vanishing LRS class II scalars $(\ca,\,\ck,\,\ce)$, for which their governing equations are
\begin{eqnarray}
\fl(\cl_N+\ca-\ck)\ca=0,\qquad (\cl_N -\ca-\frac 12\,\ck)\ck=0\qquad\mbox{and}\qquad \ce=\ca\,\ck.\label{covscheqns}
\end{eqnarray}

\section{First-Order Perturbations}

We introduce a perturbation operator which is denoted $\de$. Subsequently, a perturbed tensor $\tilde V_{\m\dots\n}$ may be approximated to first-order according to,
\begin{eqnarray}
\tilde V_{\m\dots\n} \approx V_{\m\dots\n} +\de V_{\m\dots\n}, 
\end{eqnarray}
where $V_{\m\dots\n}$ is the background tensor, and $\de V_{\m\dots\n} $ is the first-order component of order $\epsilon$, where $|\epsilon| << 1$.

It is important to stress that a background three-tensor will have a corresponding first-order tensor which is not spacelike with respect to the background. Instead, it will attain a first-order timelike component \cite{Battye1995,Burston,Mukohyama2000}. Similarly, a background two-tensor will have a corresponding first-order tensor which is not a two-tensor with respect to the background. 

Consider some perturbed two-tensor which therefore satisfies the covariant relationships $\tilde U_{\al} \tilde  N^\al=0$ and $\tilde U_{\al} \tilde  n^\al=0$. These relationships, which also hold true in the background spacetime $(U_{\al} N^\al=0$ and $U_{\al} n^\al=0)$, are then expressed to first-order according to
\begin{eqnarray}
U_\al \,\de N^\al + N^\al\,\de U_\al=0\qquad\mbox{and}\qquad U_\al \,\de n^\al + n^\al\,\de U_\al=0 \label{fospacal}.
\end{eqnarray}
Thus, it is clear that the first-order term ($\de U_\m$) is not a two-tensor with respect to the background (i.e. does not vanish when contracted with $n^\m$ or $N^\m$), and instead it now satisfies the first-order relationships \eref{fospacal}.

Now consider the perturbed two-tensor, $\tilde U_{\m\dots\n}$. This may be expanded to first-order, and its corresponding first-order component is decomposed into 1+1+2 form in the usual fashion,
\begin{eqnarray}
\tilde U_{\m\dots\n} &\approx& U_{\m\dots\n}+ \de U_{\m\dots\n}\nonumber\\
&\approx& U_{\m\dots\n}+ \de U_{\bar\m\dots\bar\n}  + (N^\al\de U_{\al\dots\n})\,N_\m  +\dots+ (N^\al\de U_{\m\dots\al})\,N_\n \nonumber\\
&&- (n^\al \de U_{\al\dots\n}) n_\m-\dots- (n^\al \de U_{\m\dots\al}) n_\n.
\end{eqnarray}
In the situation where the background two-tensor ($U_{\m\dots\n}$) vanishes, the first-order terms reduce to two-tensors according to \eref{fospacal}. We then reuse the background symbol by replacing $\de U_{\bar\m\dots\bar\n}$ with $U_{\m\dots\n}$, and similarly for scalars.

With the above definitions and conventions, the first-order geometric and Weyl quantities are,
\begin{eqnarray}
\fl\{K,\Si,\Omega,\xi,\Si_\m ,\Si_{\m\n},\Omega_\m,\ca_\m,\ca_{\m\n},\al_\m,\hat N_\m,\ce_\m,\ce_{\m\n},\cb,\cb_\m,\cb_{\m\n}\} =\mc{O}(\epsilon).
\end{eqnarray}
This set of first-order variables is gauge-invariant under infinitesimal coordinate transformations. This is a consequence of the Stewart-Walker-Sachs lemma \cite{Sachs,Stewart1974}, which states that a quantity which vanishes in the background has a corresponding first-order quantity that is gauge-invariant.
There are three quantities which do not vanish in the background, and will experience first-order increments according to
\begin{eqnarray}
\tilde \ck \approx \ck+\de \ck, \qquad \tilde \ca\approx \ca+\de\ca \qquad\mbox{and}\qquad \tilde \ce \approx \ce +\de \ce.
\end{eqnarray}
Thus $\de\ck$, $\de\ca$ and $\de\ce$ are not gauge-invariant. Finally, the first-order energy-momentum perturbations are all gauge-invariant, as they vanish on the background, hence 
\begin{eqnarray}
\{\rho,\, P,\, \mc{J},\, \mc{J}_\m,\, \Pi, \,\Pi_\m,\, \Pi_{\m\n}\}=\mc{O}(\epsilon) .
\end{eqnarray}

\section{1+1+2 Identities}

We now consider the Bianchi identities. It is the twice contracted Bianchi identities which provide the linear equations governing the energy-momentum perturbations. It is the once-contracted Bianchi identities for which the covariant, decoupled BP equations are naturally constructed with energy-momentum sources. Finally, the Ricci identities for both $n^\m$ and $N^\m$ provide the remaining covariant equations governing the full dynamics of the perturbed spacetime.

\subsection{Twice Contracted Bianchi Identities}

The twice contracted Bianchi indentities give rise to the equations governing the conservation of energy-momentum, $\na^\al T_{\m\al}=0$. They may be decomposed into 1+1+2 form to give evolution and propagation equations\footnote{\eref{121cont} derives from  $n^\m\na^\al T_{\m\al}=0$, \eref{evfrojdsa} from $N^\m \na^\al T_{\m\al}=0$ and \eref{112eulve} from  $\na^\al T_{\bar\m\al}=0$.}, which after linearization, become
\begin{eqnarray}
\
\mc{L}_n\, \rh +(\mc{L}_N+2\,\ca-\ck\,)\mc{J} +d^\al\mc{J}_\al=0,\label{121cont}\\
\mc{L}_n \mc{J} +(\cl_N +\ca-\frac32\,\ck)\Pi +d^\al \Pi_\al +(\cl_N +\ca)P+\ca\,\rh = 0,\label{evfrojdsa}\\
\mc{L}_n \mc{J}_{\bar\m} + (\cl_N-\ck)\Pi_{\bar\m}+d^\al \Pi_{\m\al}-\frac 12\, d_\m \Pi+d_\m\,P= 0.
\label{112eulve}
\end{eqnarray}
These equations (along with additional equations of state) govern the first-order 1+1+2 energy-momentum quantities. Furthermore, these equations clearly decouple from the remaining Bianchi and Ricci identities, as there are no first-order geometric quantities present. Thus, since the background quantities are prescribed, the system \eref{121cont}-\eref{112eulve} forms an initial-value problem and can be completely solved prior to solving for the remaining perturbations. These energy-momentum perturbations therefore become known sources in the following identities.

\subsection{Once Contracted Bianchi Identities}

The once-contracted Bianchi identities for Einstein's equations have a rich correspondence with the equations governing electromagnetism. We show how two decoupled, covariant and complex equations arise naturally from these identities. Therefore, we give them extra attention and provide a discussion of the fully non-linear equations before writing down the linearized system.  The identities can be expressed in terms of both the Weyl conformal curvature and energy-momentum tensor ($T_{\m\n}$), and we define
\begin{eqnarray}
B_{\n\si\ta} := \na^\al C_{\al\n\si\ta} -  (\na_{[\si} T_{\ta]\n} +\frac 13 g _{\n[\si} \na_{\ta]}T)=0\label{bianchids}.
\end{eqnarray} 
The 1+3 equations governing the GEM tensors can be found by an irreducible decomposition of \eref{bianchids}, which ultimately gives rise to two constraint and two evolution equations\footnote{The constraint \eref{divegratdas} is derived from $\p n^\al  n^\be  B_{\al\be\m}=0$ or $\p {\epsilon_{[\m}}^{\al\be}\,B _{\n]\al\be}=0$ and \eref{divb1+3} from ${\epsilon_\m}^{\be\ga} n^\al B_{\al\be\ga}=0$ or  $ \qquad \p n^\al B_{[\m\n]\al}=0$. The evolution equations, \eref{evfoe13} and \eref{evoforb13}, derive from $\p n^\al B_{(\m\n)\al}=0$ and $\p {\epsilon_{(\m}}^{\al\be}\,B _{\n)\al\be}=0$ respectively.},

\begin{eqnarray}
 D^\al E_{\m\al}+\epsilon_{\m\al\be} B^{\al\ga}{A_\ga}^\be +3\,
{B_\m}^\al \, \omega_\al \nonumber\\=\frac13\, D_\m \rho-\frac 12\, D^\al \pie_{\m\al} +\frac 13\, K \, j_\m-\frac 12\, {A_\m}^\al j_\al+\frac 32\, {\epsilon_\m}^{\al\be} j_\al \omega_\be,\label{divegratdas}\\
 D^\al B_{\m\al}-\epsilon_{\m\al\be} E^{\al\ga}{A_\ga}^\be -3\,
{E_\m}^\al \, \omega_\al\nonumber\\=(\rho+P)\,\omega_\m +\frac 12 {\epsilon_\m}^{\al\be} D_\al j_\be +\frac 12\, \epsilon_{\m\al\be} \pie^{\al\ga}{A^\be}_\ga-\frac 12\, {\pie_\m}^\al \omega_\al,\label{divb1+3}
\end{eqnarray}
\begin{eqnarray}
\fl(\mc{L}_n -\frac 13\, K )E_{<\m\n>}+\epsilon_{\al\be(\m} (D^\al+2\,\nd^\al){B_{\n)}}^\be+5\,{A_{<\m}}^\al E_{\n>\al}-{\epsilon_{(\m}}^{\al\be} E_{\n)\al}\omega_\be\nonumber\\
 =\frac 12\,(\rh+P)\, A_{\m\n}-\frac 12\,(D_{<\m} +2\, \nd_{<\m}) j_{\n>} \nonumber\\
-\frac 12\,(\mc{L}_n+\frac 13\, K)\pie_{<\m\n>}
-\frac 12\,\pie_{\al<\m}{A_{\n>}}^\al +\frac 12 {\epsilon_{(\m}}^{\al\be} \pie_{\n)\al} \omega_\be\label{evfoe13}
\end{eqnarray}
and
\begin{eqnarray}
\fl(\mc{L}_n-\frac 13\,K) B_{<\m\n>}-\epsilon_{\al\be(\m} (D^\al+2\,\nd^\al){E_{\n)}}^\be+5\,{A_{<\m}}^\al B_{\n>\al} -{\epsilon_{(\m}}^{\al\be} B_{\n)\al}\omega_\be\nonumber\\
=\frac12\, \epsilon_{\al\be(\m}{A_{\n)}}^\al j^\be -\frac 12\, \epsilon_{\al\be(\m} D^\al  {\pie_{\n)}}^\be-\frac 32\, j_{<\m} \omega_{\n>}.
\label{evoforb13}
\end{eqnarray}
In the absence of energy-momentum sources, it is easy to inspect that these fully non-linear 3+1 Bianchi identities \eref{divegratdas}-\eref{evoforb13} are invariant under the simultaneous transformation $E_{\m\n} \rightarrow B_{\m\n}$ and $B_{\m\n} \rightarrow -E_{\m\n}$. Therefore, following the work of \cite{BurstonEM1}, a natural way to decouple the system is to construct a complex-conjugate pair of the form $E_{\m\n} \pm\rmi\, B_{\m\n}$. It is then straightforward to reintroduce the energy-momentum quantities into the calculations.

We now proceed to write the linearized Bianchi identities in a 1+1+2 form. Therefore, we define three first-order 1+1+2 complex tensors, 
\begin{eqnarray}
\fl\Phi_{\m\n}:= \de\ce_{\bar\m\bar\n} +\rmi\, \de\cb_{\bar\m\bar\n},\qquad \Phi_{\m}:= \de\ce_{\bar\m} +\rmi\, \de\cb_{\bar\m}\qquad\mbox{and}\qquad \de\Phi := \de\ce+\rmi\,\de\cb,\label{comphidefs}
\end{eqnarray}
where $\Phi_{\m\n}$ and $\Phi_\m$ are gauge-invariant, whereas $\de\Phi$ is not because $\de \ce$ because in the background $\Phi = \ce \ne0$. Furthermore, when constructing these complex quantities, it is found that some of the other geometric quantities also naturally combine, and thus we make three additional definitions
\begin{eqnarray}
\Lambda_{\m\n}:= \de \Si_{\bar\m\bar\n} +\rmi\, {\epsilon^\al}_{(\m} \de \ca_{\bar\n)\al} ,\qquad\Lambda_{\m}:= \de \hat N_{\bar\m} +\rmi\,({\epsilon_\m}^\al\,\de \Si_{\al}+\de\Omega_{\bar\m}), \\
\Upsilon_{\m}:= \de \al_{\bar\m} +\rmi\, {\epsilon^\al}_{\m} \de \ca_{\al}.
\end{eqnarray}

With these definitions, \eref{divegratdas}-\eref{evoforb13}, under a first-order perturbation, and in 1+1+2 form are\footnote{These equations are derived as follows: \eref{drofphi} from $(N^\ga n^\al n^\be+\rmi \,\epsilon^{\be\ga} n^\al) B_{\al\be\ga} $=0, \eref{drofphimu} from $ n^\al n^\be B_{\bar\m\be\ga} +\rmi\,{\epsilon_{\bar\m}}^{\be\ga} n^\al  B_{\al\be\ga}=0$, \eref{evforphi} from $ N^\be N^\ga n^\al B_{\be\ga\al} +\rmi\,N^\ga \epsilon^{\al\be} B_{\ga\al\be}=0$, \eref{evforphimu} from $(N^\n n^\al B_{(\bar\m\n)\al} +\rmi\,N^\n \,{\epsilon_{(\bar\m}}^{\al\be} B_{\n)\al\be}=0$ and \eref{evoforphimu} from $n^\al B_{(\bar\m\bar\n)\al} +\rmi\,{\epsilon_{(\bar\m}}^{\al\be} B_{\bar\n)\al\be}=0$.},
\begin{eqnarray}
\de[(\cl_N-\frac 32 \ck) \Phi]+ d^\al  \Phi_\al -3\,\rmi\,\ce\,\Omega=\mc{G},\label{drofphi}\\
(\cl_N-\ck)\Phi_{\bar\m}+ d^\al\Phi_{\m\al} -\frac12\,\de(d_{\bar\m} \Phi) +\frac 32\,\ce\,\Lambda_\m=\mc{G}_{\m},\label{drofphimu}\\
\de(\cl_n\Phi)-\rmi\,\epsilon^{\al\be} d_\al \Phi_\be+\frac 32\,\ce\,(\Si-\frac 23\,K)  -\rmi\,2\,\xi=\mc{F},\label{evforphi}\\
\cl_n\Phi_{\bar\m}+ \frac32\,\ce\,\Upsilon_{\m}-\rmi\, {\epsilon_{(\m}}^{\al} d^\ga \Phi_{\ga)\al}-\rmi\, \frac 12\,{\epsilon_{\m}}^\al\,\left[ \de(d_\al\Phi)  -(2\,\ca+\ck) \Phi_\al\right]  =\mc{F}_\m,\label{evforphimu}\\
\cl_n\Phi_{\bar \m\bar\n}
+\rmi\,{\epsilon_{(\m}}^\al (\cl_N+\frac 12\ck+2\,\ca) \Phi_{\bar\n)\al} -\rmi\,{\epsilon_{\{\m}}^\al d_{|\al|}\Phi_{\n\}}-\frac32\ce\,\Lambda_{\m\n}=\mc{F}_{\m\n}, \label{evoforphimu}
\end{eqnarray}
where the vertical bars in \eref{evoforphimu} imply that the index is excluded in the symmetric brackets. The first-order energy-momentum, or forcing, terms have been defined
\begin{eqnarray}
\fl \mc{G}:=\frac1 3 \,\cl_N(\rho-\frac 32\,\Pi)+\frac 34\,\ck\,\Pi -\frac 12\,d^\al \Pi_\al+\rmi\,\frac 12\, \epsilon^{\al\be} d_\al \mc{J}_\be \label{emsourone},\\
\fl \mc{G}_\m := \frac1 3 \,d_\m\rho+\frac14\,d_\m  \Pi  -\frac 12\,(\cl_N-\ck)\Pi_{\bar\m}-\frac 12\,d^\al \Pi_{\m\al} -\rmi\,\frac12\, {\epsilon_\m}^\al( \cl_N \mc{J}_\al - d_\al\mc{J}) ,\\
\fl\mc{F} :=  -\frac 12\,\cl_n\Pi-\frac 13\,(\mc{L}_N+2\,\ca+\frac 12\,\ck) \mc{J} +\frac16\,d^\al \mc{J}_\al-\rmi\, \frac 12\,\epsilon^{\al\be}\,d_\al \Pi_\be,\\
\fl \mc{F}_\m :=  -\frac12\,\cl_n \Pi_{\bar\m}-\frac 14\, (\cl_N+2\,\ca+2\,\ck)\mc{J}_{\bar\m}  - \frac34\,d_\m\mc{J}\nonumber\\
 -\rmi\,\frac 12 \, {\epsilon_\m}^\al\,\left(\frac1 3 \,d_\al\rho+2\,d_\al  \Pi  -(\cl_N-\frac12\,\ck)\Pi_\al\right),\\
\fl\mc{F}_{\m\n} := -\frac 12\,\cl_n \Pi_{\bar \m\bar\n}-\frac 12\, d_{\{\m} \mc{J}_{\n\}} +\rmi\,\frac12\,\left[{\epsilon_{(\m}}^\al (\cl_N +\frac 12 \ck)\Pi_{\bar\n)\al}  -{\epsilon_{\{\m}}^\al d_{|\al|}\Pi_{\n\}} \right].\label{emsourcefive}
\end{eqnarray}
Each individual term in the 1+1+2 Bianchi identities \eref{drofphi}-\eref{evoforphimu} has been carefully grouped such that they remain gauge-invariant under infinitesimal coordinate transformations. For example, in \eref{evforphi} the first term, $\de(\cl_n \Phi)$, is gauge-invariant because its corresponding background value vanishes, i.e. $\cl_n \Phi=0$. 

The energy-momentum sources \eref{emsourone}-\eref{emsourcefive} appear as large complicated terms. However, the dynamics of all the energy-momentum terms can be fully prescribed using the conservation of energy-momentum equations \eref{121cont}-\eref{112eulve}.

\subsection{1+1+2 Ricci Identities}

The remaining 1+1+2 equations arise from the irreducible decomposition of the Ricci identities for both $n^\m$ and $N^\m$, which are defined respectively
\begin{eqnarray}
Q_{\m\n\si} := 2\,\na_{[\m} \na_{\n]} n_\si - R_{\m\n\si\ta} n^\ta =0,\label{ricciforlittlenm}\\
R_{\m\n\si} := 2 \na_{[\m} \na_{\n]} N_\si - R_{\m\n\si\ta} N^\ta=0.\label{riccifornm}
\end{eqnarray}

It is also useful to note that the Riemann tensor may always be expressed as a sum of its trace ($R_{\m\n}$ and $R$) and trace-free parts $(C_{\m\n\si\ta})$. Then, by substituting the Einstein equations to eliminate both the Ricci tensor and Ricci scalar in terms of the energy-momentum tensor, it becomes
\begin{eqnarray}
 R_{\m\n\si\ta}= C_{\m\n\si\ta} + g_{\m[\si} T_{\ta]\n} + g_{\n[\si} T_{\ta]\m} -\frac23\, g_{\m[\si} g_{\ta]\n} {T_\al}^\al .\label{rimantrace}
\end{eqnarray}
We now present the first-order system and note that the following equations \eref{condsa}-\eref{evforsiandk} generalize those presented in \cite{Clarkson} to include energy-momentum terms, for which we separate these into four groups.

\subsubsection{Constraints for the two-slices} These are derived as follows: \eref{condsa} derives from  $N^\m n^\si R_{\m\bar\n\si}=0$, \eref{contractfromdsa} from $\epsilon^{\m\n} n^\si R_{\m\n\si}=0$  and \eref{consttwo} from $\cs^{\n\si} R_{\bar\m\n\si}=0$.

\begin{eqnarray}
\fl-d_\m(\Si-\frac 23\, K)+2\,{\epsilon_\m}^\al d_\al \Omega -2\, d^\al \Si_{\m\al} + \ck\,\Si_\m +\ck\,{\epsilon_\m}^\al \Omega_\al -2\, {\epsilon_\m}^\al \cb_\al=\frac12\,\mc{J}_\m,\label{condsa}\\
d^\al \Omega_\al - \epsilon^{\al\be} d_\al \Si_\be -(2\,\ca+\ck) \Omega +\cb=0,\label{contractfromdsa}\\
-\frac 12\,\de(d_\m \ck)-{\epsilon_\m}^\al d_\al \xi + d^\al \ca_{\m\al} + \ce_\m=-\frac 12\,\Pi_\m. \label{consttwo}
\end{eqnarray}

\subsubsection{Constraints for the three-slices} These are derived as follows: \eref{propforsiandk} derives from $N^\m \cs^{\n\ta} Q_{\m\n\si}=0$, \eref{rotdivdecomp} from $\epsilon^{\m\n\si} Q_{\m\n\si}=0$,  \eref{propforxi} from  $ N^\m R_{\m[\bar\n\bar\si]}=0$, \eref{propforck} from $\cs^{\n\si} N^\m R_{\m\n\si}=0$, \eref{simaht765} from $n^\m n^\si R_{\m\bar\n\si}=0$, \eref{propforcamn} from $N^\m R_{\m\{\n\si\}}=0$ and \eref{propufosimmu} from $N^\m R_{\m(\bar\n\bar\si)}=0$.

\begin{eqnarray}
\cl_N(\Si-\frac 23  K)  -\frac32\,\ck\,\Si +d^\al \Si_\al -\epsilon^{\al\be}d_\al \Omega_\be=\mc{J},\label{propforsiandk}\\
(\cl_N -\ca-\ck)\Omega + d^\al \Omega_\al =0,\label{rotdivdecomp}\\
(\cl_N-\ck)\xi -\frac 12 \, \epsilon^{\al\be} d_\al \hat N_\be=0,\label{propforxi}\\
  \de[(\cl_N-\ca-\frac 12\ck)\ck]-\de(\ce-\ca\,\ck) +d^\al \hat N_\al=\frac 23\,\rho+\frac 12\,\Pi,  \label{propforck} \\
\fl\mc{L}_N(\Si_{\bar\m}+{\epsilon_{\bar\m}}^\al \Omega_\al) -\ck\,\Si_\m +d^\al\Si_{\m\al}-\frac 12\, d_\m( \Si+\frac 43\,K)-{\epsilon_\m}^\al d_\al \Omega +2\,\ca\, {\epsilon_\m}^\al \Omega_\al=0 ,\label{simaht765}\\
 \cl_N \ca_{\m\n}+ d_{\{\m} \hat N_{\n\}}  -\ce_{\m\n}=\frac 12\, \Pi_{\m\n}, \label{propforcamn}\\
(\cl_N +\frac 12 \ck)\Si_{\m\n} -d_{\{\m}(\Si_{\n\}}-{\epsilon_{\n\}}}^\al \Omega_\al)-{\epsilon_{(\m}}^\al \cb_{\n)\al}=0 \label{propufosimmu}.
\end{eqnarray}

\subsubsection{Transportation equations} These are derived as follows: \eref{Raychadsa} from  $\p^{\m\si} Q_{\m\n\si} n^\n =0$,  \eref{omegaln} from ${\epsilon_{\bar\m}}^{\al\ga} R_{\al\be\ga} n^\be=0$ and \eref{evpropforNhat} from $n^\al N^\be R_{\al\be\bar\m}=0$.
\begin{eqnarray}
\mc{L}_n K+\de[(\cl_N+\ca-\ck)\ca]   + d^\al \ca_\al\label{kdot}=\frac12\,(\rh+3\,P),\label{Raychadsa}\\
{\epsilon_\m}^\al\mc{L}_n\Omega_\al-\frac 12 \, \cl_N \ca_{\bar\m} +\frac 12\,\de(d_\m\ca)-\frac12\,\ca\,\hat N_{\m} =0,\label{omegaln}\\
 -\cl_n \hat N_{\bar \m}+(\cl_N  +\ca)\al_{\bar\m} -(\ca+\frac 12\ck) (\Si_\m-{\epsilon_\m}^\al\Omega_\al) -{\epsilon_\m}^\al\cb_\al=\frac 12\, \mc{J}_\m\label{evpropforNhat}.
\end{eqnarray}
\subsubsection{Evolution equations} These are derived as follows: \eref{evforsimommu} $ n^\m N^\si Q_{\m(\bar \n \si)}=0$, \eref{omegadot} from $n^\m\epsilon^{\n\si} Q_{\m\n\si}=0$, \eref{evoforamn} from $n^\m R_{\m\{\n\si\}}=0$, \eref{evuforsigmn} from $ n^\m Q_{\m(\n\si)}=0$, \eref{evolforckdas}  from $n^\m \cs^{\n\si} R_{\m\n\si} =0$, \eref{evforxi} from  $n^\m \epsilon^{\n\si} R_{\m\n\si} =0$ and \eref{evforsiandk} from $n^\m N^\n n^\si R_{\m\n\si}=0$.

\begin{eqnarray}
\cl_n(\Si_{\bar\m}+{\epsilon_\m}^\al\Omega_{\bar\m} )+\de(d_\m\ca)+(\ca+\frac12\, \ck) \ca_{\m}-\ce_{\m} =-\frac 12\, \Pi_{\m}\label{evforsimommu},\\
\mc{L}_n\Omega -\frac 12\,\epsilon^{\al\be} d_\al\ca_\be -\ca\,\xi=0,\label{omegadot}\\
 \cl_n \ca_{\m\n} -(\ca+\frac 12\, \ck)\Si_{\m\n}+ d_{\{\m }\al_{\n\}} +{\epsilon_{(\m}}^\al \cb_{\n)\al}=0 \label{evoforamn},\\
\cl_n \Si_{\bar\m\bar\n} - \ce_{\m\n}-\ca\,\ca_{\m\n} +d_{\{\m} \ca_{\n\}}=-\frac 12\, \Pi_{\m\n},\label{evuforsigmn}\\
\de(\cl_n \ck)+(\ca+\frac 12 \,\ck) \,(\Si-\frac 23 \,K)  +d^\ga\al_\ga =-\mc{J}\label{evolforckdas},\\
\cl_n \xi -(\ca+\frac12\,\ck) \Omega-\frac 12\,\epsilon^{\al\be} d_\al \al_\be +\frac 12\,\cb =0,\label{evforxi}\\
\cl_n(\Si-\frac 23\, K)- d^\al \ca_\al-\de(\ce-\ca\,\ck)=-\frac13 \,(\rho+3\,P)-\frac 12\,\Pi\label{evforsiandk} .
\end{eqnarray}

For the coming decoupling section, we require evolution and propagation equations for the complex tensor $\Lambda_{\m\n}$. For the propagation equation we simply take the appropriate combination of \eref{propforcamn} and  \eref{propufosimmu}. Similarly for the evolution equation we use both \eref{evoforamn} and \eref{evuforsigmn}. They are respectively
\begin{eqnarray}
-\cl_N \Lambda_{\bar\m\bar\n}-\rmi\,{\epsilon_{(\m}}^\al \,\Phi_{\n)\al} -\frac 12\,\ck\,\Sigma_{\m\n}+\rmi\, {\epsilon_{\{\m}}^\al\,d_{\n\}} \Lambda_\al=  \rmi\,\frac 12 \,{\epsilon_{(\m}}^\al \Pi_{\n)\al},\label{ecvolsiandcamucom}\\
-\cl_n \Lambda_{\bar\m\bar\n} + \Phi_{\m\n}-\rmi\,\ca\, {\epsilon_{(\m}}^\al\,\Lambda_{\n)\al} -\rmi\,\frac 12\,\ck\, {\epsilon_{(\m}}^\al \Sigma_{\n)\al}+\rmi\, {\epsilon_{\{\m}}^\al d_{\n\}} \Upsilon_\al=\frac 12\, \Pi_{\m\n}.\label{evorlammn}
\end{eqnarray} 

\section{Decoupling the Bianchi Identities}

With the once contracted Bianchi identities expressed in complex form, it is now possible to derive a decoupled equation for $\Phi_{\m\n}$. Before, proceeding with the derivation, it is necessary to first present the commutation relationships for the three derivative operators when acting on any covariant first-order tensor $\varphi_{\m\dots\n}$,
\begin{eqnarray}
\cl_n d_{\bar\m}\varphi_{\n\dots\si} - d_\m \cl_n \varphi_{\n\dots\si}=0,\\
\cl_N d_{\bar\m}\varphi_{\bar\n\dots\bar\si}  -d_\m \cl_N\varphi_{\bar\n\dots\bar\si} =0,\\
(\cl_N+ \ca) \cl_n\varphi_{\bar\m\dots\bar\n}-\cl_n \cl_N \varphi_{\bar\m\dots\bar\n}=0.
\end{eqnarray}

 The calculation is complicated, lengthy and begins by taking the Lie derivative with respect to $n^\m$ of \eref{evoforphimu}. Then commute the two Lie derivatives, which are acting on $\Phi_{\m\n}$, and substitute \eref{evoforphimu} again to eliminate $\cl_n \Phi_{\bar\m\bar\n}$. Also commute $\cl_n d_\m$ acting on $\Phi_\m$ and substitute \eref{evforphimu} to eliminate $\cl_n \Phi_\m$. Then similarly commute $\cl_N d_\m$ and then substitute \eref{drofphimu} to eliminate $\cl_N \Phi_\m$. Throughout this entire procedure, collate any terms which have a background gravito-electric scalar ($\ce$). Finally, this collated term must be reduced using both \eref{ecvolsiandcamucom} and \eref{evorlammn}. This results in a covariant, and gauge-invariant, wave-like equation
\begin{eqnarray}
\fl\left[\cl_n \cl_n -(\cl_N+\ca-\ck\,) \cl_N  -d^\al d_\al-V\right]\Phi_{\bar\m\bar\n}+\rmi\,(4\,\ca+2\,\ck) {\epsilon_{(\m}}^\al \cl_n \Phi_{\bar\n)\al} =\mc{M}_{\m\n} .\label{tenweforphimn}
\end{eqnarray}
The potential and source have beed defined respectively,
\begin{eqnarray}
 V:=-2\, K_g-\frac12\,\ck^2-4\,\ca^2 +2 \,\ce,\\
\fl\mc{M}_{\m\n}:= \cl_n \mc{F}_{\m\n} +\rmi\, {\epsilon_{(\m}}^\al (\cl_N-\ca-\frac 32\, \ck) \mc{F}_{\n)\al} -\rmi\,{\epsilon_{\{\m}}^\al d_{|\al|} \mc{F}_{\n\}} +d_{\{\m}  \mc{G}_{\n\}}.
\end{eqnarray}
Here, $K_g$ is the Gaussian curvature of the background two-surface, i.e. $K_g:=\frac12\,^{(2)}R$, where $^{(2)}R$ is the Ricci scalar for the background two-surface.

We now proceed to further decompose the wave-like equation  for $\Phi_{\m\n}$ \eref{tenweforphimn}  into two decoupled wave-like equations for spin-weighted scalars. Both $\Phi_{\m\n}$ and $\mc{M}_{\m\n}$ are symmetric and trace-free and they are decomposed using \eref{antisyshedec},
\begin{eqnarray}
\Phi_{\m\n} =X\, \bar m_\m \bar m_\n + Y \,m_\m m_\n ,\label{phoidecmsd}\\
\mc{M}_{\m\n} =Q\, \bar m_\m \bar m_\n + P \,m_\m m_\n .
\end{eqnarray}
Here, the scalars $X$ and $P$ have spin-weight $s=2$, whereas $Y$ and $P$
  have spin-weight $s=-2$. Thus, from the decomposition of \eref{tenweforphimn}, we get two {\it decoupled, covariant and gauge-invariant, wave-like equations} for the spin-weighted scalars $X$ and $Y$,
\begin{eqnarray}
(\cl_n +4\,\ca+2\,\ck)\cl_n X -(\cl_N+\ca-3\,\ck)\cl_N X  - (d^\al+4\, \chi^\al) d_\al X\nonumber\\
-\left[-2\, K_g+\ck^2-4\,\ca^2 +4\,\chi\right]X =Q, \label{wavelikeone}
\end{eqnarray}
and
\begin{eqnarray}
(\cl_n -4\,\ca-2\,\ck)\cl_n Y-(\cl_N+\ca-3\,\ck)\cl_N Y- (d^\al-4\, \chi^\al) d_\al Y\nonumber\\
-\left[-2\, K_g+\ck^2-4\,\ca^2 +4\,\chi \right]Y=P, \label{waveliketwo}
\end{eqnarray}
where $\chi := (d^\al m^\be) (d_\al \bar m_\be)=\bar \chi$. These wave-like equations correspond to the BP equations \cite{Bardeen} for gravitational perturbations\footnote{These equations have been checked with the vacuum equations in the literature \cite{Bardeen,Chandra,Fernandes}, by using a specific choice of the definitions for the null and complex conjugate-vectors as defined in \cite{Chandra}.}. These covariant and gauge-invariant wave-like equations may be solved for the spin-weighted scalars ($X$ and $Y$) and then reconstructed back into $\Phi_{\m\n}$ which is zero weighted. Finally, the real and imaginary parts give respectively $\de \ce_{\m\n}$ and $\de \cb_{\m\n}$. 

\subsection{Static Schwarzschild Coordinates}

The purpose of this section is to express the wave-like equations \eref{wavelikeone} and \eref{waveliketwo} in a pragmatic coordinate form suitable for both analytic and numerical computation. The covariant equations governing the Schwarzschild spacetime \eref{covscheqns} are solved for the background LRS class II scalars \cite{Clarkson} to give
\begin{eqnarray}
\fl\ca=\frac M{r^2}\left(1-\frac{2\, M}{r}\right)^{-\frac 12},\qquad \ck=-\frac 2 r \sqrt{1-\frac {2\, M} r}\qquad\mbox{and}\qquad\ce=-\frac {2\, M}{r^3}.
\end{eqnarray}
This corresponds to the standard static Schwarzschild line element ($ds$) using coordinates $x^\m:= (t,r,\theta,\phi)$,
\begin{eqnarray}
ds^2= -\left(1-\frac{2\,M}r\right)dt^2 +\frac 1 {\left(1-\frac {2\,M}r\right)} dr^2 + r^2\left(d\theta^2+\sin^2\theta d\phi^2\right),
\end{eqnarray}
where $M$ is the Schwazschild mass.  Without loss of generality the Cauchy time function is set equal to $t$ and the lapse function is therefore,
\begin{eqnarray}
\al = \sqrt{1-\frac{2\,M}r}.
\end{eqnarray}
Furthermore,  the complex-conjugate vectors defining the two-metric $(\cs_{\m\n})$ is chosen such that
\begin{eqnarray}
m^\m =\frac 1{\sqrt{2}\, r }\left(0,0,1,\rmi\, \mbox{cosec}\,\theta\right)\label{ccvector},
\end{eqnarray}
and the Gaussian curvature of the two-spheres and other necessary quantities reduce to 
\begin{eqnarray}
\fl K_g=\frac1{r^2},\qquad\si=1,\qquad \chi_\m = [0,0,0,\rmi\,\cos\theta]\qquad\mbox{and}\qquad\chi=\frac{\cot^2\theta}{r^2}.
\end{eqnarray}
The shift-vectors ($\be^\m$ and $b^\m$) vanish and the Lie derivatives \eref{1p4lide} and \eref{cldecomsa} when operating on scalars reduce to 
\begin{eqnarray}
\cl_n =\frac 1\al \, \frac\pl{\pl t} \qquad\mbox{and} \qquad \cl_N = \al \,\frac\pl{\pl r} .
\end{eqnarray}
Therefore, the wave-like equations \eref{wavelikeone}-\eref{waveliketwo} become
\begin{eqnarray}
\fl\left[-\frac 1{\al^2}  \, \frac{\pl^2}{\pl t^2} -\left(\frac{12\,M-4\,r}{r^2-2\,M\,r}\right)  \, \frac\pl{\pl t}\right]X\nonumber\\
 \fl+\left[\al^2\,\frac{\pl^2}{\pl r^2}-\frac{(10\,M-6\,r)}{r^2}\frac\pl{\pl r}+\frac 2{r^2}\left(\frac{r^2-6\, M\,r +6\, M^2}{r^2-2\,M\,r}\right)\right]X\nonumber\\
\fl + \frac1{r^2}\left[\frac1{\sin\theta} \frac\pl {\pl \theta}\left(\sin\theta\, \frac\pl {\pl \theta}\right) +\frac 1{\sin^2\theta} \frac {\pl ^2}{\pl\phi^2} +\rmi\,\frac{4\,\cos\theta}{\sin^2\theta}\frac{\pl}{\pl\phi} +4\,\cot^2\theta \right]X=-Q\label{compwlone}
\end{eqnarray}
and
\begin{eqnarray}
\fl\left[-\frac 1{\al^2}  \, \frac{\pl^2}{\pl t^2} +\left(\frac{12\,M-4\,r}{r^2-2\,M\,r}\right)  \, \frac\pl{\pl t}\right]Y\nonumber\\
\fl +\left[\al^2\,\frac{\pl^2}{\pl r^2}-\frac{(10\,M-6\,r)}{r^2}\frac\pl{\pl r}+\frac 2{r^2}\left(\frac{r^2-6\, M\,r +6\, M^2}{r^2-2\,M\,r}\right)\right]Y\nonumber\\
\fl + \frac1{r^2}\left[\frac1{\sin\theta} \frac\pl {\pl \theta}\left(\sin\theta\, \frac\pl {\pl \theta}\right) +\frac 1{\sin^2\theta} \frac {\pl ^2}{\pl\phi^2} -\rmi\,\frac{4\,\cos\theta}{\sin^2\theta}\frac{\pl}{\pl\phi} +4\,\cot^2\theta \right]Y=-P\label{compwltwo}.
\end{eqnarray}

These wave-like equations  \eref{compwlone} and \eref{compwltwo} give the exciting prospect of calculating the first-order gravitational fields resulting from a first-order energy-momentum perturbation. They constitute an initial-value problem, whereby the energy-momentum is specified as a perturbation to an initial spacelike three-slice as the source to resultant gravitational fields. In particular, they are suitable for modelling gravitational radiation.

An initial hypersurface may be constructed at $t=t_0$ and the energy-momentum perturbations are specified
\begin{eqnarray}
\{\rho, P, \mc{J},\mc{J}_\m, \Pi,\Pi_\m,\Pi_{\m\n}\}|_{t=t_0}.
\end{eqnarray}
The conservation of energy-momentum equations \eref{121cont}-\eref{112eulve}, plus equations of state, may then be used to evolve the energy-momentum perturbations for all time. In the case of numerical computations, they can be integrated up until the events of interest.

Thus, with the energy-momentum sources ($Q$ and $P$) now known, we further require initial conditions for $X$ and $Y$, i.e.
\begin{eqnarray}
X|_{t=t_0}, \qquad \frac{\pl X}{\pl t}\Big|_{t=t_0}\qquad\mbox{and}\qquad Y|_{t=t_0} ,\qquad \frac{\pl Y}{\pl t}\Big|_{t=t_0}.
\end{eqnarray}
Since we know $m^\m$ and $\bar m^\m$, this corresponds to initial Cauchy data given by
\begin{eqnarray}
\Phi_{\m\n}|_{t=t_0}\qquad\mbox{and} \qquad \frac{\pl \Phi_{\m\n}}{\pl t}\Big|_{t=t_0}.
\end{eqnarray}
To obtain this initial data, one must solve the non-trivial first-order initial-value problem. That is to say, one must solve the first-order system arising from the Bianchi and Ricci identities \eref{condsa}-\eref{evforsiandk} on the initial spacelike three-slice. It is also important to separate the particular solutions arising from the energy-momentum perturbations from the homogeneous solutions which exist irrespectively.

Finally, in the absence of energy-momentum sources ($P=Q=0$), the homogenous solutions to \eref{compwlone} and \eref{compwltwo} encompass the usual vacuum perturbations, whereby the vacuum operators are separable (as there is a common factor of $1/r^2$ out the front of all angular terms) and spin-weighted spherical harmonic solutions arise \cite{Penrose}.

\section{Summary and Discussion}

We have shown how constructing complex quantities for the GEM tensors was successful for finding decoupled, gauge-invariant and covariant, wave-like equations  \eref{compwlone} and \eref{compwltwo}. These equations form an initial-value problem and are suitable for calculating first-order gravitational perturbations which are induced by first-order energy-momentum sources. 

The Weyl conformal curvature tensor comprises ten trace-free components of the Riemann tensor.  The two spin-weighted {\it complex} scalars $X$ and $Y$ correspond to four of the ten components. That is, $X$ and $Y$ correspond to $\Phi_{\m\n}$  through \eref{phoidecmsd} which further corresponds to $\ce_{\m\n}$ and $\cb_{\m\n}$ through \eref{comphidefs}. Now since the two-tensors $\ce_{\m\n}$ and $\cb_{\m\n}$ are trace-free with respect to the two-surface, they each have two components for a total of four.

The study of the remaining six components is the subject of future research.  Immediate difficulties arise when one attempts to derive wave-like equations for the remaining complex GEM tensors. Consider $\Phi_\m$, by taking the Lie derivative with respect to $n^\m$ of \eref{evforphimu}, one is confronted with a term involving $\cl_n \Upsilon_{\bar\m}$, which does not arise naturally. Finally, $\de \Phi$ is not gauge-invariant and thus the physical interpretation of this scalar is nebulous.

\end{document}